\documentclass[prb,twocolumn,showpacs,reprint,preprintnumbers,amsmath,amssymb]{revtex4-1}

\usepackage{color}
\usepackage{graphicx}
\usepackage{dcolumn}
\usepackage{bm}

\preprint{submitted to Physical Review B - Rapid Communication}

\begin{document}

\title{Generalized Slater-Pauling rule for the inverse Heusler compounds}

\author{S. Skaftouros$^1$}
\author{K. \"{O}zdo\u{g}an$^2$}\email{kozdogan@yildiz.edu.tr}
\author{E. \c{S}a\c{s}{\i}o\u{g}lu$^{3,4}$}\email{e.sasioglu@fz-juelich.de}
\author{I. Galanakis$^1$}\email{galanakis@upatras.gr$^1$}
\affiliation{$^1$Department of Materials Science, School of
Natural Sciences, University of Patras,  GR-26504 Patra,
Greece\\
$^2$Department of Physics, Yildiz Technical University, 34210
\.{I}stanbul, Turkey \\ $^3$Peter Gr\"{u}nberg Institut and
Institute for Advanced Simulation, Forschungszentrum J\"{u}lich
and JARA, 52425 J\"{u}lich, Germany\\ $^4$ Department of Physics,
Fatih University, 34500, B\"{u}y\"{u}k\c{c}ekmece, \.{I}stanbul,
Turkey}

\begin{abstract}
We present extensive first-principles calculations on the inverse
full-Heusler compounds having the chemical formula X$_2$YZ where
(X\,=\,Sc, Ti, V, Cr or Mn), (Z\,=\,Al, Si or As) and the Y ranges from
Ti to Zn. Several of these alloys are identified to be
half-metallic magnets. We show that the appearance of
half-metallicity is associated in all cases to a Slater-Pauling
behavior of the total spin-magnetic moment. There are three
different variants of this rule for the inverse Heusler alloys
depending on the chemical type of the constituent transition-metal
atoms. Simple arguments regarding the hybridization of the
\emph{d}-orbitals of neighboring atoms can explain these rules. We
expect our results to trigger further experimental interest on
this type of half-metallic Heusler compounds.
\end{abstract}


\maketitle

The rise of nanotechnology and nanoscience during the last decade
brought to the center of scientific research new phenomena and
materials. Spintronics and magnetoelectronics compose one of the
most rapidly expanding field in nanoscience.\cite{ReviewSpin}
Half-metallic magnetic compounds play a crucial role in this
development.\cite{ReviewHM} These materials present usual metallic
behavior for the one spin direction while an energy gap in the
band structure is present in the other spin direction similarly to
semiconductors.\cite{ReviewGala06,ReviewGala07} The possibility of
creating 100\% spin-polarized current has triggered the interest
on such compounds.\cite{FelserRev} De Groot and collaborators in
1983 have initially suggested based on electronic structure
calculations that NiMnSb, a semi-Heusler alloy, is a
half-metal\cite{deGroot} and since then several half-metallic
compounds have been discovered.\cite{Pickett07} Several aspects
concerning the implementation of half-metallic alloys in realistic
devices, like tunnelling magnetic junctions or giant
magnetoresistive junctions and spin-injectors, have been discussed
in literature.\cite{Chadov11,Lezaic06,Mavropoulos05}

The family of Heusler alloys incorporates more than 1000 members
almost all crystalizing in a close-packed cubic structure similar
to the binary semiconductors.\cite{Graf11} Most of them are metals
exhibiting diverse magnetic phenomena. The lattice is a f.c.c.
with four equidistant  sites as basis along the diagonal of the
unit cell. \cite{ReviewGala06} There are two families of Heusler
alloys. The semi-(or half-)Heuslers have the chemical formula XYZ
where the sequence of the sites is X-Y-void-Z. The X and Y are
transition-metal elements and Z is a \textit{sp}-element and the
structure is known as the C1$_b$ lattice. The second subfamily
consists the full-Heusler compounds with the chemical formula
X$_2$YZ. When the valence of the X is larger than Y, the atomic
sequence is X-Y-X-Z and the structure is the well known $L2_1$ one
with prototype Cu$_2$MnAl.\cite{Galanakis09} When the valence of
the Y elements is the largest, the compounds crystallize in the
so-called \emph{XA} structure, where the sequence of the atoms is
X-X-Y-Z and the prototype is Hg$_2$TiCu.\cite{Galanakis09} The
latter alloys are also known as inverse Heusler compounds. Several
inverse Heuslers have been studied using first-principles
electronic structure calculations in
literature.\cite{Galanakis09,Liu08,Meinert11,Li09,Luo08,Xu11,Pugaczowa12,Kervan}
In all cases the \emph{XA} structure is energetically preferred to
the $L2_1$ structure. The latter has been also confirmed by
experiments on Mn$_2$CoGa and Mn$_2$CoSn films as well as Co doped
Mn$_3$Ga samples,\cite{Winterlik11,Meinert11b,Klaer11,Alijani12}
but experiments on Mn$_2$NiSb revealed that the actual arrangement
of the atoms at the various sites can be influenced by the
preparing method.\cite{Luo09} Inverse Heuslers are interesting for
applications since they combine coherent growth on semiconductors
with large Curie temperatures which can exceed the 1000 K as in
the case of Cr$_2$CoGa.\cite{APL11}

\begin{figure*}
\includegraphics[scale=0.55]{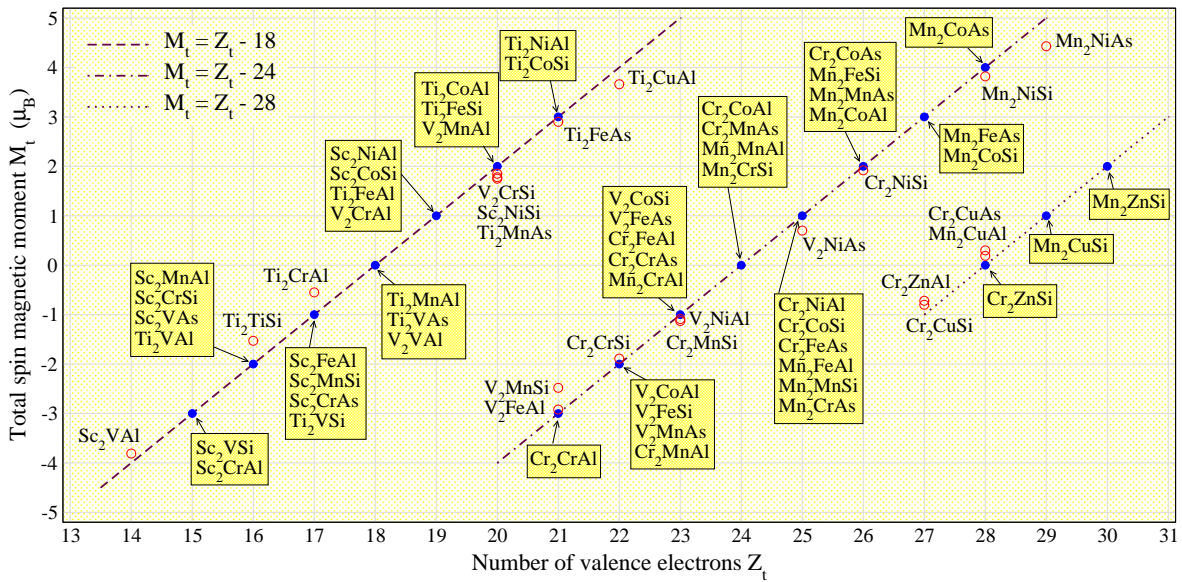}
\vskip -0.2 cm \caption{(color online). Total spin magnetic
moments per unit cell (in $\mu_B$) as a function of the total
number of valence electrons in the unit cell for several compounds
under study. The lines represent the three different forms of the
Slater-Pauling rule. The compounds within the shaded frames follow
one of these rules and are perfect half-metals, while the rest of
the alloys slightly deviate and we denote their total spin
magnetic moment with an open red circle. Notice that the sign of
the spin magnetic moments has been chosen so that the
half-metallic gap is in the spin-down band.} \label{fig1}
\end{figure*}

Slater and Pauling had shown in two pioneering papers that in the
case of binary magnetic alloys when we add one valence electron in
the compound this occupies spin-down states only and the total
spin magnetic moment decreases by about 1
$\mu_B$.\cite{Slater,Pauling} Interestingly a similar behavior can
be also found in half-metallic Heusler alloys. It was shown that
in the case of the semi-Heusler compounds like NiMnSb the total
spin magnetic in the unit cell, $M_t$ scales, as a function of the
total number of valence electrons, $Z_t$, following the relation
$M_t=Z_t-18$,\cite{Galanakis02} while in the case of the $L2_1$
full-Heuslers this relation becomes
$M_t=Z_t-24$.\cite{Galanakis02b} These  Slater-Pauling (SP) rules
connect the electronic properties (appearance of the half-metallic
behavior) directly to the magnetic properties (total spin magnetic
moments) and thus offer a powerful tool to the study of
half-metallic Heusler compounds. It has been shown that also
quaternary or quinternary half-metallic Heusler alloys obey the SP
rule,\cite{Galanakis04,Quaternary,Quinternary} and a generalized
version exists when we pass from a half-metallic semi-Heusler to a
half-metallic full-Heusler alloy.\cite{Galanakis12} The aim of the
present communication is to exploit whether a generalized version
of the SP rule can be also extracted for the inverse half-metallic
full-Heusler alloys where all sites obey the tetrahedral symmetry.
Such a relation would offer an extra strong tool in the study of
half-metals.

For our calculations we used the full-potential nonorthogonal
local-orbital minimum-basis band structure scheme
(FPLO)\cite{Koepernik} within the  generalized gradient
approximation (GGA)\cite{GGA} to study the electronic and magnetic
properties of all the inverse X$_2$YZ alloys where X\,=\,Sc, Ti,
V, Cr or Mn,  Y is a transition-metal atom ranging from Ti to Zn,
and Z\,=\,Al, Si or As. First, we determined the equilibrium
lattice constants using total energy calculations and a dense
20$\times$20$\times$20 $\mathbf{k}$-mesh grid to carry out the
numerical integrations, and at the equilibrium constant we
calculated the electronic and magnetic properties with more
accuracy.  In the following we use the superscripts A and B to
distinguish the two X atoms sitting at the two inequivalent sites
in the \emph{XA}-structure (see Fig.\,1 in
Ref.\,\onlinecite{Galanakis09} for a schematic representation of
the structure). The X$^A$ atom has the same local environment in
the crystal as the Y atom.

\begin{figure*}
\includegraphics[scale=0.25]{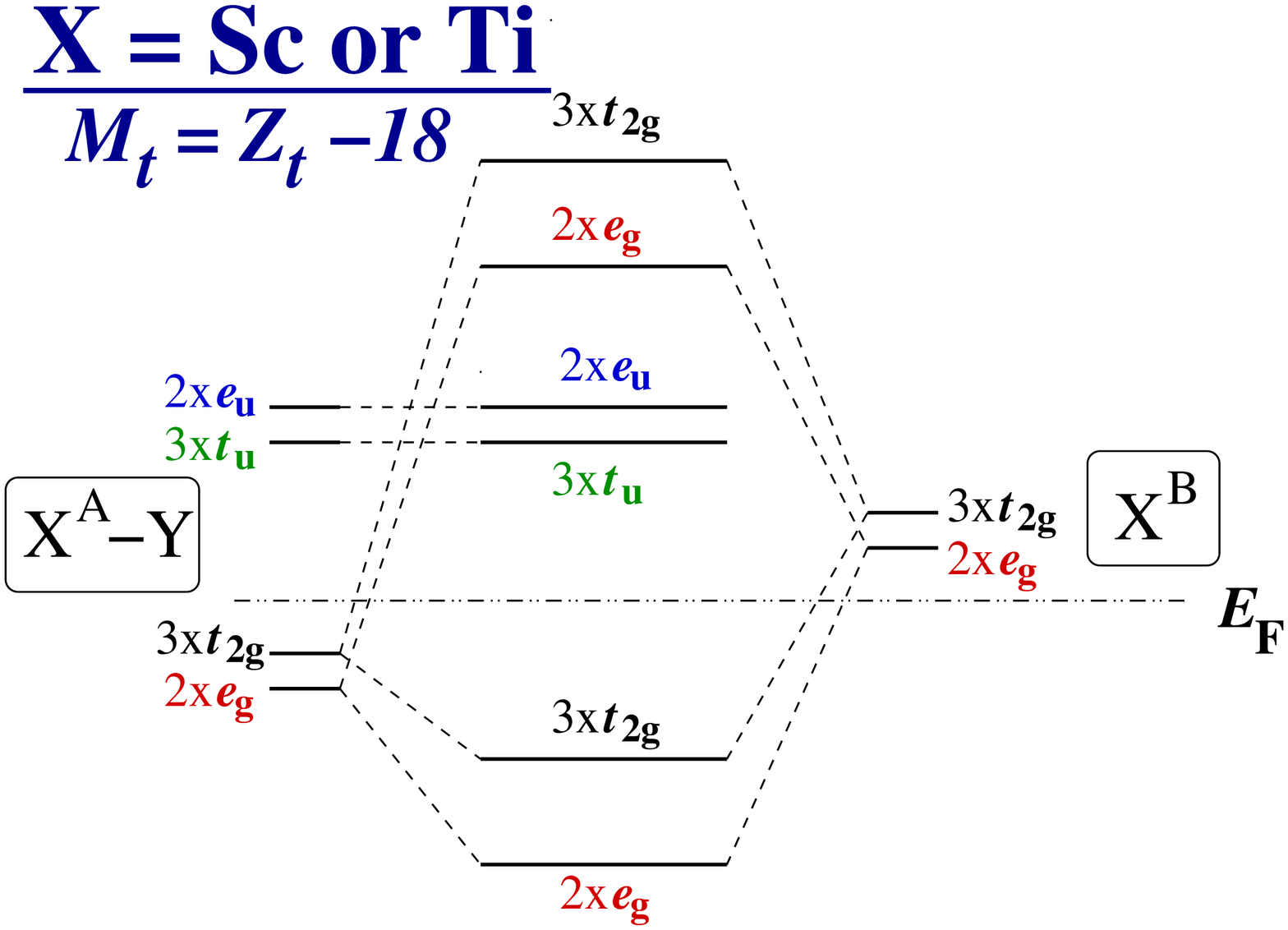} \hskip 0.1cm
\includegraphics[scale=0.25]{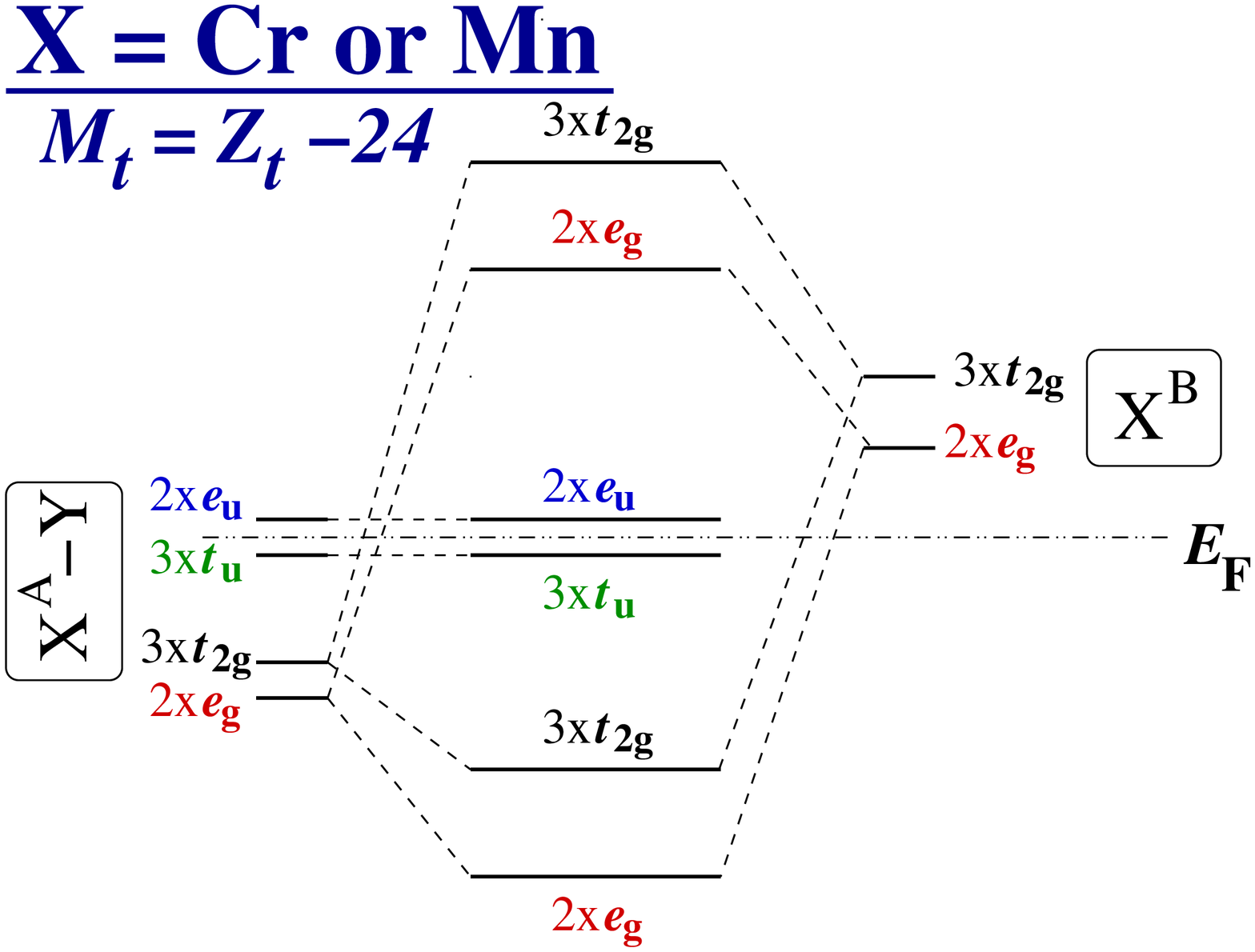} \hskip 0.1cm
\includegraphics[scale=0.25]{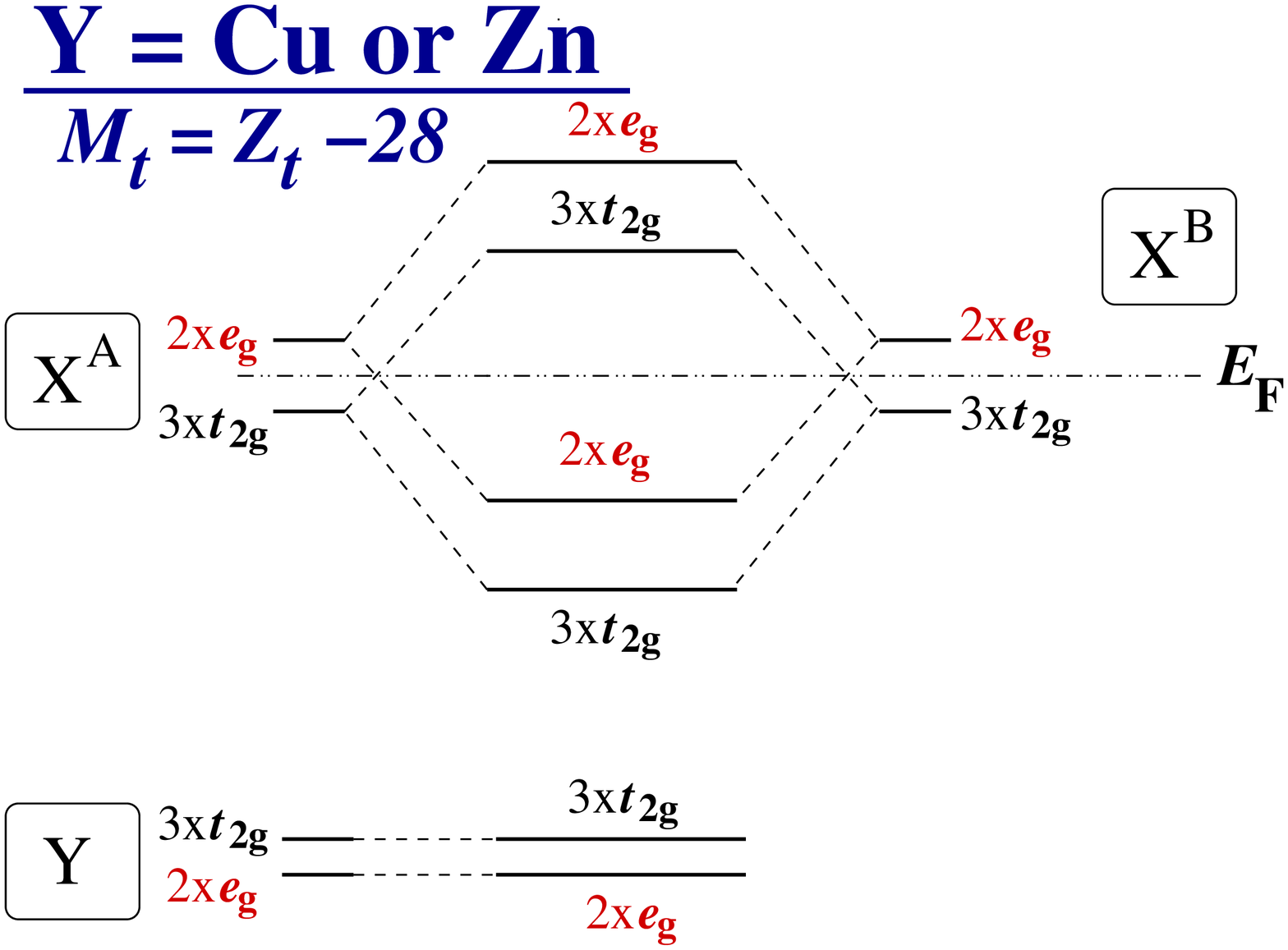}
\vskip -0.2 cm \caption{(color online). Possible hybridizations
between spin-down \emph{d}-orbitals  sitting at different sites in
the case of the inverse X$_2$YZ Heusler compounds.  The names of
the orbitals and the superscripts follow the nomenclature
discussed in the text; the coefficient represents the degeneracy
of each orbital. Note that in the spin-down band structure, there
are also one \emph{s} and three \emph{p} occupied states due to
the Z atom. With black color we denote the 3x$t_{2g}$ orbitals,
red the 2x$e_g$, blue the 2x$e_u$ and green the 3x$e_{1u}$ ones.}
\label{fig2}
\end{figure*}

In Fig.\,\ref{fig1} we have plotted the total spin magnetic moment
versus the total number of valence electrons in the unit cell for
all studied compounds which were found to be half-metals. We have
considered in all cases that the half-metallic gap in the density
of states (DOS) is located in the spin-down band. In the cases of
negative total spin magnetic moments in the figure the spin-up are
the minority and the spin-down the majority states. . Our results
can be grouped along three lines representing three variants of
the SP rule. Along the $M_t=Z_t-18$ line we find the Sc and Ti
based alloys, along the $M_t=Z_t-24$ line the alloys with X\,=\,Cr
or Mn, and finally along the $M_t=Z_t-28$ line the compounds where
Y is Cu or Zn. The alloys with X\,=\,V are dispersed between the
first two lines.  A very interesting consequence of the SP rules
are the Heuslers compounds with a zero value of their total spin
magnetic moment which are made of magnetic constituents and which
are belong to a special class of half-metallic antiferromagnets
also known as fully-compensated ferrimagnets  (we have not
included the semiconducting or the simple-metallic systems in Fig.
\ref{fig1}).\cite{Leuken,Wurmehl06,Galanakis07}

Prior to the discussion of the SP rules in the inverse Heusler
alloys, we should shortly discuss the origin of the rule in the
half-metallic $L2_1$ full-Heuslers (for more details see
Ref.\,\onlinecite{Galanakis02b}). In their case the corresponding
SP rule is $M_t=Z_t-24$. The role of the \textit{sp} element is to
provide in the spin-down electronic band structure a single
\textit{s} and a triple-degenerated \textit{p} band deep in
energy; they are located below the \emph{d}-states and accommodate
\emph{d}-charge from the transition metal atoms. Due to the more
complex \emph{d}-\emph{d} hybridizations in these alloys with
respect to the semi-Heuslers one has first to consider the
interaction between the X elements. Although the symmetry of the
$L2_1$ lattice is the tetrahedral one, the X elements themselves,
if we neglect the Y and Z atoms, form a simple cubic lattice and
sit at sites of octahedral symmetry.\cite{Galanakis02b}  The
\emph{d}-orbitals of the neighboring X atoms hybridize creating
five bonding \emph{d}-states, which after hybridize with the
\emph{d}-orbitals of the Y atoms creating five occupied and five
unoccupied \emph{d}-hybrids, and five non-bonding \emph{d}-hybrids
of octahedral symmetry (the triple-degenerated $t_{1u}$ and
double-degenerated $e_u$ states). These non-bonding hybrids cannot
couple with the orbitals of the neighboring atoms, since they do
not obey the tetrahedral symmetry, and only the $t_{1u}$ are
occupied leading to a total of 12 occupied spin-down
states.\cite{Galanakis02b} In the case of semi-Heuslers, like
NiMnSb, the situation is simpler. The \emph{d}-orbitals of the two
transition metal atoms hybridize strongly creating five occupied
bonding and five unoccupied antibonding \emph{d}-states in the
spin-down band structure.\cite{Galanakis02} As a result there are
in total exactly nine occupied spin-down states and the SP
relation is $M_t=Z_t-18$.\cite{Galanakis02}

All half-metallic Sc- and Ti-based alloys follow the $M_t=Z_t-18$
rule as for the semi-Heusler compounds but its origin is
different. A close look at the atom-resolved spin-moments, DOS and
band structure can give more information on the origin of this
rule. The Sc$^A$ (Ti$^A$) and the Y atoms sit at sites of the same
symmetry and their \emph{d}-orbitals hybridize as in the usual
$L2_1$ full-Heuslers creating five bonding \emph{d}-hybrids and
five non-bonding. The five Sc$^A$-Y (Ti$^A$-Y) bonding
\emph{d}-hybrids in their turn hybridize with the
\emph{d}-orbitals of the Sc$^B$ (Ti$^B$) atoms creating again
bonding and antibonding states. The difference with the $L2_1$
full-Heuslers is that the Sc$^A$ (Ti$^A$) and Y atoms have a large
energy separation of their \emph{d}-orbitals and as a result the
five Sc$^A$-Y (Ti$^A$-Y) non-bonding \emph{d}-hybrids, the
$t_{1u}$ and $e_u$ states are very high in energy and they are
empty while in the usual $L2_1$ full-Heuslers the
triple-degenerated $t_{1u}$ states where occupied (see Fig.
\ref{fig2} for a schematic representation of the \emph{d}-\emph{d}
hybridizations in all cases under study). Thus now there are 9 and
not 12 occupied states in the spin-down band and the SP rule is
$M_t=Z_t-18$ instead of $M_t=Z_t-24$. When the X atom is Cr or Mn
the energy position of the \emph{d}-states of the Cr$^A$ (Mn$^A$)
and Y atoms is much closer and the non-bonding spin-down $t_{1u}$
states are occupied like in usual full-Heuslers (see middle panel
in Fig.\,\ref{fig2}), and the SP rule for them is the $M_t=Z_t-24$
one. The case where X is V is more complex since the V is in
between the Sc (Ti) and Cr (Mn) transition-metal elements. As a
result no general rule can be deduced and the behavior of the
total spin magnetic moment of the V$_2$YZ compounds is material
specific, \textit{e.g.} V$_2$MnAl follows the $M_t=Z_t-18$ SP rule
while V$_2$MnSi is close to the $M_t=Z_t-24$ SP rule.

\begin{figure}
\includegraphics[scale=0.55]{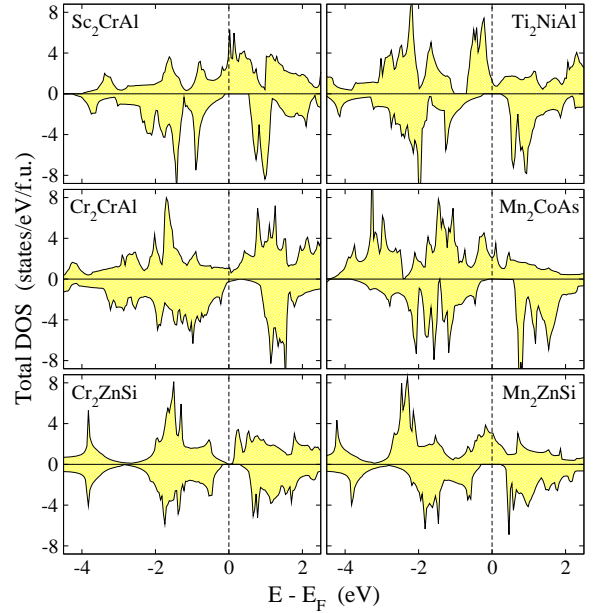}
\vskip -0.2 cm \caption{(color online). Total density of states
(DOS) per formula unit, which coincides with the per unit cell
value, for some selected compounds under study. Positive
(negative) DOS values correspond to the spin-up (spin-down) states.
We plot the DOS so that the half-metallic gap is  located in the
spin-down band structure. The zero of the
energy axis corresponds to the Fermi level.} \label{fig3}
\end{figure}

In Fig. \ref{fig3} we present the total DOS for some selected
cases and in Table \ref{table1} the calculated equilibrium lattice
constant and the spin magnetic moments. In the case of the Sc- and
Ti-based compounds the theoretical lattice constants exceed the 6
\AA\ due to the large extension of the Sc and Ti
\emph{d}-wavefunctions while for the V-, Cr- and Mn-based
compounds it is around 5.9 \AA\ when Z\,=\,Al, around 5.8 \AA\ when
Z\,=\,As and around 5.7 \AA\ when Z\,=\,Si. The total spin magnetic
moment varies from almost -4 $\mu_B$ for Sc$_2$VAl up to almost 4
$\mu_B$ for Ti$_2$CuAl as shown in Fig.\,\ref{fig1} for the Sc- and
Ti-based alloys. In the case of the Sc-based compounds the Sc
atoms carry small spin magnetic moments of around 0.5 $\mu_B$ and
mainly the Y atoms carry the spin magnetic moment as shown in
Table\,\ref{table1} for two selected cases: Sc$_2$CrAl and
Sc$_2$MnAl. When we move to the Ti-based compounds, the Ti atoms
contrary to the Sc ones can carry significant spin moment which
can even reach a value of almost 2 $\mu_B$ as in Ti$_2$NiAl. This
is due to the larger number of valence electrons of the Ti atoms.
In the case of the V-(Cr- or Mn-) based alloys all X transition
metal atoms possess significant values of spin magnetic moments.
Concerning the atoms at the Y sites: when Y\,=\,V, Cr or Mn they carry
spin moments with absolute values larger than 2 $\mu_B$, when Y\,=\,Ni
the Ni-spin moment is close to zero, and when Y\,=\,Fe or Co the
absolute value of the Fe (Co) spin magnetic moment can vary from 0
to $\sim$1 depending on the material. The Z atoms carry negligible
spin magnetic moments. The relative orientation of the spin
magnetic moments is in most cases dictated by the Bethe-Slater
curve which states that most transition metal atoms tend to have
parallel spin magnetic moments with the exception of Mn and Cr
which tend to have antiparallel spin magnetic moment with respect
to their nearest neighbors.\cite{Bethe-Slater}

\begin{table}
\caption{For selected compounds we present the calculated
equilibrium lattice constant (2nd column),  and the atomic and
total spin magnetic moments (in $\mu_B$) (3rd-7th columns). We use
the symbols A and B to denote the two early transition metal atoms
sitting at different sites (see text for explanation). Note that
the total spin magnetic moment is given per unit cell. $Z_t$ is
the total number of valence electrons in the unit cell and the
last column presents the form of the SP rule obeyed by
the compound.}
\begin{ruledtabular}
\begin{tabular}{llrrrrrrc}
X$_2$YZ & a(\AA ) & $m^{X(A)}$ & $m^{X(B)}$ & $m^Y$ & $m^Z$ &
$m^{t}$ & $Z_t$ & SP \\ \hline Sc$_2$CrAl & 6.68 & 0.38 & -0.02 &
-3.57 & 0.21 & -3 & 15 &
$Z_t$-18 \\
Sc$_2$MnAl & 6.58 & 0.59 & 0.34 & -3.04 & 0.11 & -2 & 16 &
$Z_t$-18 \\
Ti$_2$FeAl & 6.14 &1.21 & 0.84 & -1.02 &  -0.02 & +1 & 19&
$Z_t$-18\\
Ti$_2$CoAl & 6.14 & 1.50 & 0.80 & -0.23 & -0.07 & +2 & 20&
$Z_t$-18\\
Ti$_2$NiAl & 6.20 & 1.84 & 1.15 & 0.10 & -0.08 &+3 & 21&
$Z_t$-18\\
V$_2$MnAl & 5.94 & 1.54 &-0.41 & 0.94 & -0.07 & +2 & 20 &
$Z_t$-18\\
V$_2$CoAl & 5.92 & -1.77 &0.28 & -0.54 & 0.03 & -2 & 22 &
$Z_t$-24\\
Cr$_2$CrAl & 5.92 & -2.47 & 1.89 & -2.47 & 0.12 & -3 & 21 &
$Z_t$-24\\
Mn$_2$CrAl & 5.85& -1.68& 2.81& -2.23& 0.09&  -1& 23&
$Z_t$-24\\
Mn$_2$MnAl & 5.78& -1.51& 2.70 & -1.51& 0.06& 0&24&
$Z_t$-24\\
Mn$_2$FeAl & 5.74& -1.85&  2.81& 0.05& -0.01& +1&25&
$Z_t$-24\\
Mn$_2$CoAs & 5.74& -0.04&  3.04& 0.98& 0.02& +4&28&
$Z_t$-24\\
Cr$_2$ZnSi & 5.85 & -1.89 & 1.93 & 0.01 & -0.05 & 0 & 28 &
$Z_t$-28\\
Mn$_2$CuSi & 5.75 & -1.92 & 2.91 & -0.01 & 0.01 & +1 & 29 &
$Z_t$-28 \\
Mn$_2$ZnSi & 5.78 & -0.80 & 2.77 &  0.01 & 0.02 & +2 & 30&
$Z_t$-28
\end{tabular}
\end{ruledtabular}
\label{table1}
\end{table}

Finally we would like to dwell on the compounds where the Y atom
is Zn or Cu. Similar compounds have been previously studied using
first-principles
calculations.\cite{Luo11,Luo12,Wei11,Wei11b,Wang11} It has been
suggested in Ref. \onlinecite{Luo12} that they should be a variant
of the SP rule: $M_t=Z_t-28$. Among the (X\,=\,Sc, Ti or V)
studied compounds in this work, only Ti$_2$CuAl showed a behavior
close to half-metallicity. In this case the compound is close to
the $M_t=Z_t-18$ line, as shown in Fig.\,\ref{fig1} and the same
arguments stand as for the Sc- and Ti-based inverse full-Heuslers
discussed above. In the case of the Cr (Mn)-based alloys
half-metallicity was accompanied in all cases by a $M_t=Z_t-28$ SP
behavior of the total spin magnetic moment. The Cu (Zn) atoms have
all their 3\emph{d}-states occupied and they form a narrow band
below the energy window shown in Fig.\,\ref{fig3} for the
Cr$_2$ZnSi and Mn$_2$ZnSi compounds. As a result also their
atom-resolved spin magnetic moments in Table \ref{table1} are
close to zero. Thus relevant for the discussion of the
half-metallic gap shown in their DOS is only the interaction
between the Cr (Mn) atoms sitting at the A and B nearest
neighboring sites. The \emph{d}-orbitals Cr$^A$ (Mn$^A$) hybridize
with the \emph{d}-orbitals of the  Cr$^B$ (Mn$^B$) atoms forming 5
occupied \emph{d}-hybrids in the spin-down band and five
unoccupied \emph{d}-hybrids (see right panel in Fig.\,\ref{fig2}
for a schematic representation of the \emph{d}-\emph{d}
hybridization scheme). This behavior is similar to the
semi-Heuslers discussed above. Thus in total in the spin-down band
we have 14 occupied states: the 5 \emph{d}-states of Cu (Zn), the
one \textit{s}- and three \textit{p}-states created by the Z atom
and the five Cr$^A$-Cr$^B$ (Mn$^A$-Mn$^B$) bonding
\emph{d}-hybrids. This explains the $M_t=Z_t-28$ SP rule for these
compounds. Finally, we have to note in the DOS presented for
Cr$_2$ZnSi in Fig.\,\ref{fig3} that in the spin-up band we do have
a metallic behavior as in usual half-metals but the conduction and
the valence bands touch each other creating a zero-width gap. Such
compounds are known as spin gapless semiconductors and they form a
special category of half-metals.\cite{Wang} There is experimental
evidence for such behavior in the inverse
Mn$_2$CoAl.\cite{Ouardi12} Among our studied alloys we have
identified six such compounds but the results will be presented in
detail elsewhere since such a study exceeds the scope of the
present Communication.\cite{Skaftouros}

In conclusion, we have presented extensive first-principles
calculations on the inverse full-Heusler compounds having the
chemical formula X$_2$YZ where (X\,=\,Sc, Ti, V, Cr or Mn),
(Z\,=\,Al, Si or As) and the Y ranges from Ti to Zn. Several of
these alloys were identified to be half-metallic magnets. We have
shown that the appearance of half-metallicity is associated in all
cases to a SP behavior of the total spin-magnetic moment. When X
is Sc or Ti, the total spin magnetic moment per formula unit (or
unit cell) in $\mu_B$ follows the rule $M_t=Z_t-18$ where $Z_t$ is
the total number of valence electrons in the unit cell. When
X\,=\,Cr- or Mn, the variant followed by $M_t$ is $M_t=Z_t-24$,
and when X\,=\,V the form of the SP rule is material specific.
Both forms of the SP rule have been explained based on simple
hybridization arguments of the transition metal \emph{d}-orbitals.
Finally we have shown that when X is Cr or Mn and Y is Cu or Zn,
the half-metallic compounds follow a $M_t=Z_t-28$ rule due to the
fully-occupied Cu (Zn) \emph{d}-orbitals. In the case of semi- and
$L2_1$ full-Heusler compounds the formulation of the SP rules
offered a theoretical basis on which the experimental design of
novel materials took place. We expect that our study and the
formulation of simple rules connecting the electronic and magnetic
properties also in the case of the inverse full-Heusler compounds
will strengthen the interest on half-metallic magnets for
spintronics and magnetoelectronics applications offering to
experimentalists a more extended theoretical basis for the design
of novel half-metallic compounds.

\end{document}